# Adaptive Load Shedding for Grid Emergency Control via Deep Reinforcement Learning

Ying Zhang, *Member, IEEE*, Meng Yue, *Member, IEEE*
Interdisciplinary Science Department
Brookhaven National Laboratory
Upton, NY, USA

Jianhui Wang, *Fellow, IEEE*
Electrical and Computer Engineering Department
Southern Methodist University
Dallas, TX, USA

*Abstract*—Emergency control, typically such as under-voltage load shedding (UVLS), is broadly used to grapple with low voltage and voltage instability issues in real-world power systems under contingencies. However, existing emergency control schemes are rule-based and cannot be adaptively applied to uncertain and floating operating conditions. This paper proposes an adaptive UVLS algorithm for emergency control via deep reinforcement learning (DRL) and expert systems. We first construct dynamic components for picturing the power system operation as the environment. The transient voltage recovery criteria, which poses time-varying requirements to UVLS, is integrated into the states and reward function to advise the learning of deep neural networks. The proposed method has no tuning issue of coefficients in reward functions, and this issue was regarded as a deficiency in the existing DRL-based algorithms. Case studies illustrate that the proposed method outperforms the traditional UVLS relay in both the timeliness and efficacy for emergency control.

## I. Introduction

Emergency control enables to minimize the chance of occurrence and impact of blackouts and thus is paramount for resilient grid operation. The root cause of most blackouts in the past decades originates from voltage instability due to severe contingencies, such as faults [1], [2]. Practical power system utilities widely adopt under-voltage load shedding (UVLS) for emergency control. For instance, the California Independent System Operator (CAISO) declared a statewide emergency and ordered utilities to shed about 1,000 megawatts on a day during the COVID-19 pandemic [1].

Emergency control can be formulated as a sequential decision-making problem to recover the normal system operation quickly, and the approaches mainly fall into three categories: 1) rule-based [3], 2) optimal power flow (OPF) or optimal control [4], [5], and 3) learning-based [6]–[11]. A traditional load shedding control scheme deployed in the energy industry is rule-based and described as "shedding loads by UVLS relay if voltages at certain nodes drop below a pre-defined threshold for some duration." This rule is designed offline based on conceived typical operation scenarios and cannot adaptively meet the requirements of real-time resilient operation when confronting uncertain and floating load consumption and renewable generation. Another control paradigm in the literature is model predictive control (MPC), where control actions are determined at fixed time snapshots by solving a constrained optimization model within a finite time horizon [5]. Nevertheless, the optimization procedure in MPC is model-based and computationally intensive, and the performance highly relies on the specific modeling of prediction and operation models.

Recently, learning-based methods for grid emergency control capture broad interests in academia. In machine learning, [6] proposes a decision tree-based control approach, while [7] applies extreme learning machines to UVLS against fault-induced delayed voltage recovery (FIDVR) events. Then, reinforcement learning (RL), such as Q-learning, provides a panel of methods that allows learning a goal-oriented control law from interactions with a system or its simulation model [9]. However, RL is inefficient in processing high dimensional observation and action spaces. Then deep RL (DRL) by integrating deep learning is developed by the research community and holds the promise to the highly nonlinear online optimization solving. [10] proposes a deep Q-network (DQN)-based UVLS scheme, and a deep deterministic policy gradient (DDPG) approach is applied in [11]. However, the coefficients in these reward functions of [10] and [11] require careful tuning, and a combination of heuristics based on prior knowledge and trial-and-error selection might be necessary. On the other hand, [11] does not involve the detailed modeling of various dynamic components in power systems, and the heterogeneity increases the intrinsic stochasticity and then the learning difficulty in DRL.

The existing works [10] and [11] directly use voltage series at several time snapshots as the inputs of neural networks (NNs) and lack a thorough picturing of temporal system dynamics under time-varying voltage recovery requirements. To address the concerns and improve the learning efficiency, we propose a novel and adaptive DRL-based framework by integrating two defined UVLS metrics. The states are defined as the voltage deviation with respect to the transient voltage recovery criteria (TVRC). The main contributions are twofold.

- We propose a set of effective reward functions in DRL that direct the NN agent to minimize the shedding loads and to obey the TVRC standard simultaneously. Moreover, the developed DRL scheme has a clear physical significance, and there is no significant issue of coefficient tuning.

- We introduce historical operation experience of the traditional rule-based UVLS relays as an expert system to the experience reply, which accelerates the DRL process. The experiences from the UVLS relay used in industrial practice are readily available. The proposed method based on the expert system could achieve real-time effective load shedding within

This work was supported by the Advanced Grid Modeling Program, Office of Electricity Delivery and Energy Reliability of the U.S. Department of Energy.

[1] Los Angeles Times. [Online]. Available:
https://www.latimes.com/california/story/2020-08-16/covid-19-restrictions-help-push-california-power-grid-to-breaking-point-amid-extreme-heat

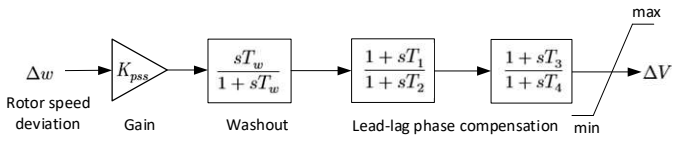

Fig.1. The diagram of PSS operation

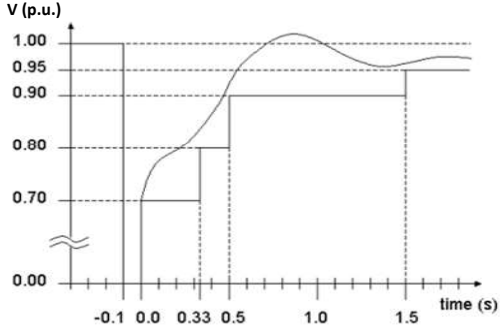

Fig.2. Transient voltage recovery criteria in power systems [14]

only milliseconds and thus outperforms the UVLS relay that often provides outdated actions.

## II. PROBLEM FORMULATION OF UVLS AGAINST FIDVR

Considering a dynamic power grid, the emergency control problem is a highly nonlinear and non-convex optimization problem over a finite time horizon, formulated as follows [5]:

$$\min_{\boldsymbol{u}} \int_{t_0}^{t_0+T} C(\boldsymbol{x}(t), \boldsymbol{y}(t), \boldsymbol{u}(t))\, dt \quad (1)$$

$$\text{s.t.} \begin{cases} \dot{\boldsymbol{x}}(t) = f(\boldsymbol{x}(t), \boldsymbol{y}(t), d(t), \boldsymbol{u}(t)) & (1a) \\ 0 = g(\boldsymbol{x}(t), \boldsymbol{y}(t), \boldsymbol{u}(t)) & (1b) \\ \underline{\boldsymbol{S}} \le G(\boldsymbol{x}(t), \boldsymbol{y}(t), \boldsymbol{u}(t)) \le \overline{\boldsymbol{S}} & (1c) \\ \underline{\boldsymbol{u}} \le \boldsymbol{u}(t) \le \overline{\boldsymbol{u}} & (1d) \end{cases}$$

where $t \in [t_0, t_0+T]$, $T$ is the control horizon time, and $t_0$ denotes the beginning time instant; $\boldsymbol{x}$ is a state vector, such as rotor angles, angular speeds, and so on; $\boldsymbol{y}(t)$ represents the algebraic state variables of the power grid, which is typically the voltages at buses of the grid; $\boldsymbol{u}(t)$ is a vector of the adopted control variables (actions), such as generator tripping or load shedding; $C(\boldsymbol{x}(t), \boldsymbol{y}(t), \boldsymbol{u}(t))$ is a cost function. During the procedure, the optimal control strategy is computed and implemented per the observation instant $T_m$.

Eqs. (1a) and (1b) represent differential-algebraic equations (DAEs) that describe power system dynamics after a given contingency or event, such as a fault. A power grid is modeled as a transmission network with dynamic components at some nodes. The network connection is formulated as follows:

$$\begin{bmatrix} \boldsymbol{G}(t) & -\boldsymbol{B}(t) \\ \boldsymbol{B}(t) & \boldsymbol{G}(t) \end{bmatrix} \begin{bmatrix} \boldsymbol{V}_x(t) \\ \boldsymbol{V}_y(t) \end{bmatrix} = \begin{bmatrix} \boldsymbol{I}_x(t) \\ \boldsymbol{I}_y(t) \end{bmatrix} \quad (2)$$

where $\boldsymbol{G}$ and $\boldsymbol{B}$ denote the system nodal conductance and susceptance matrices, and $G_{ij} + jB_{ij}$ is the nodal admittance of nodes $i$ and $j$; $V_{xi}(t) + jV_{yi}(t)$ and $I_{xi}(t) + jI_{yi}(t)$ are the complex voltage and current injection on node $i$ at time $t$.

Then, dynamic components are equivalent to current sources in (2) on the corresponding attached buses and could be synchronous machines, excitation systems, turbine governors, and power system stabilizers (PSSs). Fig.1 depicts the operation diagram of a PSS as an example. Limited by the space, the modeling details of these specific components are omitted here and can be found in [12]. Eq. (1c) represent the constraints of $\boldsymbol{x}(t)$ or $\boldsymbol{y}(t)$ during the control horizon, and for example, $\underline{\boldsymbol{S}}$ and $\overline{\boldsymbol{S}}$ could denote the lower and upper bounds of the load consumption and renewable generation, which might impact the dynamic system operation largely [13]. Eq. (1d) indicates the constraints of control action variables during the horizon. The DAEs in (1) are solved by a numerical iteration method at each integration step, specifically, the Newton-Raphson and the predictor-corrector methods [4].

Effective UVLS for emergency control is to solve the optimization model (1) with the cost function as follows at each conduction time step $t$, expressed as

$$C(\boldsymbol{x}(t), \boldsymbol{y}(t), \boldsymbol{u}(t)) = \sum_{i=1}^{n} u_{Li}(t) P_{Li}(t) \quad (3)$$

where $n$ is the number of load buses that could be controlled, and $P_{Li}(t)$ is the initial load at $t$ on the $i$th load node; $u_{Li}(t)$ is the percentage of loads to shed and $\boldsymbol{u}(t) = \{u_{Li}(t)\}$.

UVLS requires considering FIDVR that reflects a phenomenon in practical power systems, where the system voltages maintain at significantly reduced levels for several seconds after a fault is cleared [14]. See Fig.2 for a typical TVRC curve. After fault clearance, this standard requires that voltages should return to at least 0.7, 0.8, 0.9, and 0.95 of their nominal values. within 0, 0.33, 0.5, and 1.5 s, respectively. Given this criterion, eq. (1c) includes these voltage constraints.

## III. PROPOSED DRL-BASED SOLUTION

### A. Markov Decision Process

The Markov decision process (MDP) for making sequential decision is defined by the tuple $(\mathcal{S}, \mathcal{A}, p, r)$, the state space $\mathcal{S}$, the action space $\mathcal{A}$, and unknown state transition probability $p$, $\mathcal{S} \times \mathcal{S} \times \mathcal{A} \to [0, \infty]$, of the next state $\boldsymbol{s}_{t+1} \in \mathcal{S}$ given the current state $\boldsymbol{s}_t \in \mathcal{S}$ and action $\boldsymbol{a}_t \in \mathcal{A}$. At each time step, the environment results in a bounded reward $r: \mathcal{S} \times \mathcal{A} \to [r_{min}, r_{max}]$ on each transition. $\rho_\pi(\boldsymbol{s}_t)$ and $\rho_\pi(\boldsymbol{s}_t, \boldsymbol{a}_t)$ are used to denote the state and state-action marginal of the trajectory distribution induced by a policy $\pi(\boldsymbol{a}_t|\boldsymbol{s}_t)$. A time step later, the new states become $\boldsymbol{s}_{t+1}$ according to $\rho_\pi(\boldsymbol{s}_t, \boldsymbol{a}_t)$.

The goal of standard MDP is to find a control policy $\pi$ that maximizes the expected sum of rewards:

$$\max_{\pi} J(\pi) = \sum_{t=0}^{T} \mathbb{E}_{(\boldsymbol{s}_t, \boldsymbol{a}_t) \sim \rho_\pi} [r_t(\boldsymbol{s}_t, \boldsymbol{a}_t)] \quad (4)$$

where $r_t(\boldsymbol{s}_t, \boldsymbol{a}_t)$ denotes the reward function, and $\mathbb{E}(\cdot)$ denotes the expectation function.

The time-series decision making in MDP could be implemented by DRL, owing to the exploration capability of NNs towards nonlinear relationships via deep learning. In DRL, the NN agent(s) observe the states from the environment and learns the control strategies via each interaction with the learning environment. By adjusting NN parameters via the interaction, the agents gradually evolve the knowledge in the direction of increasing the reward.

### B. UVLS Design for DRL

The proposed DRL-based load shedding scheme following TVRC is shown in Fig.3. With the states as the input and the control actions as output, the NNs work as the agents are trained

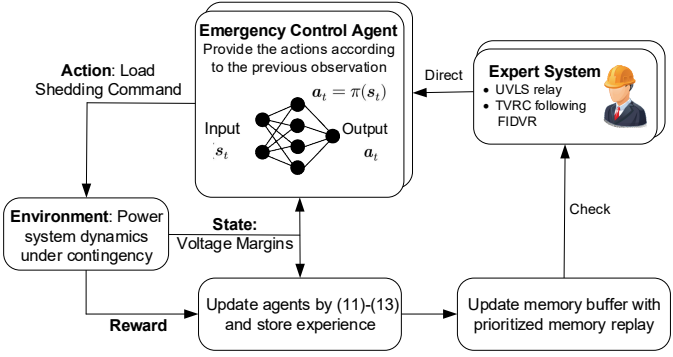

Fig.3. The proposed DRL-based UVLS paradigm

## Algorithm: DRL Training for UVLS

1. **Input**: System dynamic operation model
2. **Initialization**: the learning rate $\alpha$, the discount rate $\gamma$, the size of replay buffer, the size of mini-batch $N_d$.
   **for** each training episode **do**
3.    Initialize agent states and actions by the $\varepsilon$-greedy policy.
      **for** $t = t_0: t_0 + T$ **do**
4.       Get the reward at $t$ by (8), and new state and store them as a transition $(s_t, a_t, r_t, s_{t+1})$ into a replay buffer.
5.       Get the current $Q$ vector at state $s_t$.
6.       Sample from the replay buffer to obtain the tuple $(s_t(i), a_t(i), r_t(i), s_{t+1}(i))$, and $i = 1, 2, \ldots, N_d$.
7.       set $\hat{Q}(s_t, a_t) = \begin{cases} r_t & \text{for terminal} \\ r_t + \gamma \max Q(s_{t+1}, a_t) & \text{otherwise} \end{cases}$
8.       Train and update the agent by performing gradient descent on (13).
      **end**
   **end**
9. **Output**: NN agent with parameters $\theta^*$.

with the aids of the historical experiences to explore effective action. For UVLS, the control actions as the decision variables are expressed at each action time step as

$$a_t = u(t) = [u_{L1}(t), u_{L2}(t), \ldots, u_{Ln}(t)] \quad (5)$$

where $a_t \in \mathbb{R}^{1 \times n}$ and we define the action at each controlled load bus as either 0 (no load shedding) or 1 (10% shedding of the initial load), i.e., $u_{Li} \in \{0, 0.1\}$ as the control variables at $t$; the control action space at each time step is discrete with a dimension of $2^n$. The total dimension of the time-series actions is $(2^n)^{N_t}$, in which $N_t$ is the number of all action time steps.

The environment (power system operation) is formulated in Eqs. (1a) and (1b). It should be noted that the voltage recovery standard is time-varying during the dynamic process after contingencies following FDVIR. To truly reflect the adjustable margin of the system operation, we select the distance of voltage magnitudes from the TVRC as the observed states of the DRL agents. These states at $t$ are denoted as $O_t = \{\Delta V_j(t)\}$ with $j \in \{1, 2, \ldots, N\}$, where $N$ is the number of the nodes under monitoring. Specifically, the voltage deviation from the TVRC at node $j$, $\Delta V_j(t)$, is calculated by

$$\Delta V_j(t) = \begin{cases} V_j(t) - 0.7 & t \in [T_{fc}, T_{fc} + 0.33) \\ V_j(t) - 0.8 & t \in [T_{fc} + 0.33, T_{fc} + 0.5) \\ V_j(t) - 0.9 & t \in [T_{fc} + 0.5, T_{fc} + 1.5) \\ V_j(t) - 0.95 & t \in [T_{fc} + 1.5, t_0 + T] \end{cases} \quad (6)$$

where $T_{fc}$ is the moment of a fault clearance.

To capture the dynamics of system states, the recent $N_r$ observations are stacked as the input of agents at time $t$:

$$s_t = [O_{t-N_r-1}, \ldots, O_{t-1}] \quad (7)$$

where $O_\tau$ denotes the observations at the previous time step $\tau$, $\tau \in \{t - N_r - 1, \ldots, t - 1\}$.

UVLS is to maintain the voltage stability following the TVRC envelope while shedding the controllable loads as less as possible. Once taking the load shedding actions $a_t$, the reward at time $t$ for UVLS is defined and calculated by:

$$r_t = \begin{cases} \sum_{j=1}^{N} \min\{\Delta V_j(t), 0\} & \text{If } \forall j, \Delta V_j(t) < 0 \quad (8a) \\ \sum_{i=1}^{n} \frac{(P_{Li}(t) - \Delta P_{Li}(t))}{P_{L0}} \times 100\% & \text{Otherwise} \quad (8b) \end{cases}$$

where the percentage of voltage deviation $\Delta V_j(t)$ with respect to 1.0 p.u. is calculated, and $\Delta P_{Li}(t)$ is the shedding amount on bus $i$; $P_{L0}$ is the initial load before the UVLS process, and $P_{Li}(t)$ is the load before taking the action at $t$; (8b) calculates the remaining loads in terms of percentage of $P_{L0}$.

The reward for episode $k$ is calculated as

$$R_k = \sum_{t \in [t_0, t_0 + T]} r_t \quad (9)$$

Note that the proposed reward when satisfying the TVRC is always non-negative and greater than that in (8a) that only provides negative values. As a result, no additional coefficients are introduced to balance the role of voltages and loads in the reward function simultaneously. This design avoids the expensive tuning issue in the multi-objective problem, compared with [10] and [11]. It guarantees that a larger reward always gives a correct and positive incentive to direct the learning of agents. Besides the reward function, we define the success rate of UVLS in each episode or test:

$$\alpha_k = \begin{cases} 0 & \text{If } \forall t \text{ and } \forall j, \ \Delta V_j(t) < 0 \\ 1 & \text{Otherwise} \end{cases} \quad (10)$$

where $\alpha_k = 0$ implies that the adopted time-series actions in the current round fail to satisfy the TVRC standard.

During the DRL procedure, one always expects a relatively large total amount of the remaining loads by summing up the rewards in (8b), and a success rate $\alpha_k = 1$.

### C. DRL Implementation with UVLS Relay Experiences

The dynamics of each state after using different control strategies are intrinsically stochastic in the grid with heterogeneous components. Random "blind" actions and early premature policy in DRL might lead to a voltage collapse, even the load flow divergence at any time step. This failure is overwhelming for the real-world grid operation and prolongs the offline training time and the number of episodes. It calls on the participation of an expert system in artificial intelligence, which is defined as "a computer program that behaves like a human expert in some useful ways" [15]. On the other hand, industry practice currently adopts UVLS relays to shed a contain amount of loads at all controlled nodes in a stepwise manner, if the monitored bus voltages fall below the TVRC. These actions are conservative and outdated but safe. Hence, to improve the learning efficiency of DRL, the rule-based UVLS relay scheme, which is manually designed by power engineers, behaves like an expert system, and their actions are stored and leveraged for NN training.

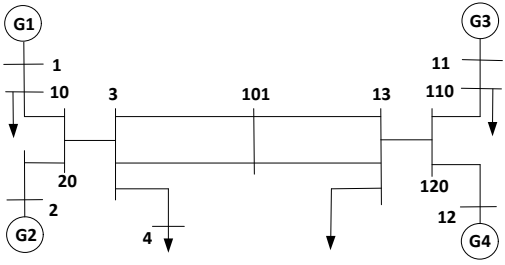

Fig. 4. The two-area four-machine power system.

Here we adopt a DQN algorithm to solve the MDP process as an example. In Q-learning, at $t$, and the agent updates the action-reward $Q$ function via the classical Bellman equation:

$$Q'(\boldsymbol{s}_t, \boldsymbol{a}_t) = Q(\boldsymbol{s}_t, \boldsymbol{a}_t) + \alpha(r_t(\boldsymbol{s}_t, \boldsymbol{a}_t) \\ + \gamma \max Q(\boldsymbol{s}_{t+1}, \boldsymbol{a}_{t+1}) - Q(\boldsymbol{s}_t, \boldsymbol{a}_t)) \quad (11)$$

where $\gamma \in [0,1]$ is a discount rate, and $\alpha$ is the learning rate.

The DQN training process adopts the replay buffer and greedy search techniques [16]. For the buffer replay, these experiences from the UVLS relay operations are stored in $\mathcal{D}_{\text{expert}}$. Let $\mathcal{D}$ be the replay buffer consisting of tuples, denoted as $\{\boldsymbol{s}_t, \boldsymbol{s}_{t+1}, \boldsymbol{a}_t, r_t\}$ that includes the expert experience $\mathcal{D}_{expert}$ and random exploration $\mathcal{D}_{rand}$. The $\varepsilon$-greedy search encourages the exploration of the action space, and as the training continues, the action selection relies more on the action policy from $Q(\boldsymbol{s}_{t+1}, \boldsymbol{a}_{t+1})$, according to:

$$\boldsymbol{a}_t = \pi(\boldsymbol{s}_t) = \begin{cases} \text{random action} & \text{if } \xi < \varepsilon_t \\ \arg\max_{\boldsymbol{a}_{t+1}} Q(\boldsymbol{s}_{t+1}, \boldsymbol{a}_{t+1}) & \text{otherwise} \end{cases} \quad (12)$$

where $\xi$ is a random number in (0,1); the searching criteria $\varepsilon_t$ is updated from the last episode by $\varepsilon_t = \varepsilon_{t-1}\eta$, and $\eta$ is a decay factor.

A Q-network is trained by minimizing the Bellman squared residual via gradient descent on NN parameters $\boldsymbol{\theta}$:

$$\mathcal{L}_Q(\boldsymbol{\theta}) = \mathbb{E}[\tfrac{1}{2}(Q(\boldsymbol{s}_t, \boldsymbol{a}_t) - \hat{Q}(\boldsymbol{s}_t, \boldsymbol{a}_t))^2] \quad (13)$$

where in theory, the target $Q$ function is

$$\hat{Q}(\boldsymbol{s}_t, \boldsymbol{a}_t) = r_t(\boldsymbol{s}_t, \boldsymbol{a}_t) + \gamma \cdot \max Q(\boldsymbol{s}_{t+1}, \boldsymbol{a}_{t+1}) \quad (14)$$

The DRL procedure in offline training is summarized in pseudocode. After then, on an online stage, a well-trained NN adaptively provides the load shedding strategies in different operating conditions.

## IV. SIMULATION RESULTS

We test the proposed algorithm on a two-area four-machine system (shown as Fig.4) and a 68-bus test system, which are widely used for dynamic research [10]. The test system contains sub-transient generators, static exciters, thermal turbine governors, and PSSs. The power system dynamic is simulated in Power System Toolbox [17], and the simulation step is 0.01s; the voltage observations are from phasor measurement units (PMUs). The simulation time of the system operation is 15 seconds. Various learning scenarios cover different combinations of initial load conditions, fault duration time, and fault locations. The fault duration time is randomly set in [0.05s, 0.07s]; for exhibiting the adaptivity for offline training and online test, the loads randomly fluctuate with 90% to 120% of the ranges, resulting in various pre-fault operating conditions [16]. The control actions are conducted on the load buses 4, 10, 13, and 110 in the two-area test system.

TABLE I
PARAMETER SETTING IN DRL

| Parameter | # | Parameter | # |
|---|---|---|---|
| Action dimensions at each $t$ | $2^4$ | The number of training episodes | 3500 |
| Size of hidden layers | {60, 30, 15} | Memory buffer size $\mathcal{D}_{expert}, \mathcal{D}_{rand}$ | 2000, 6000 |
| $\alpha$ | 0.0001 | Size of batch memory | 2000 |
| $\gamma$ | 0.95 | $\eta$ | 0.999 |

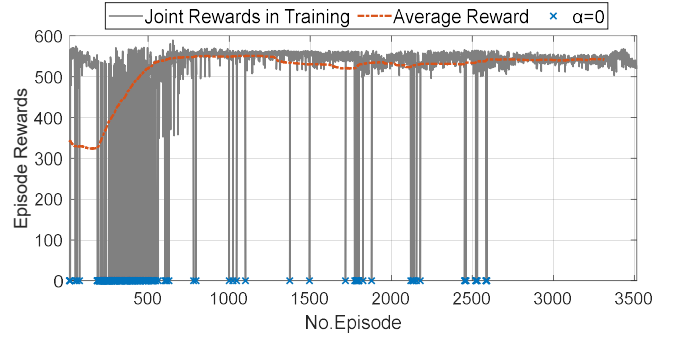

Fig. 5. Learning process during the offline training in the two-area test system. The episodes with $\alpha = 0$ are marked on the grey line.

The NNs have three fully connected layers and use rectified linear unit (ReLU) activation functions in the hidden layers. Table I lists the adopted hyperparameters. The agent provides the shedding actions per second. The recent ten sets of observations are stacked and used as the input of NNs with a dimension of 120. After the training, we test the trained DRL agent 2000 times for online tests. All the tests are performed in a 3.20 GHz, 32 GB of RAM, Intel Core i7 computer.

### A. Learning Process and Computational Efficiency

Fig. 5 illustrates the DRL process in offline training and exhibits the joint rewards by multiplying $R_k$ and $\alpha_k$ at each episode. The positive values in this curve denote the sum of the percentage of the remaining loads at all shedding steps. We analyze the performance by calculating the average rewards in successive 500 episodes. It can be observed that the average rewards gradually increase, while the frequency of $\alpha = 0$ in (10), i.e., a failure of the action decision that the agents advise, is lower and lower. Finally, the agents evolve their strategies to grapple with different operating conditions and contingencies.

Numerical experiments in online tests are carried out to investigate the computational efficiency of the proposed algorithm. The average decision time for 2000 online tests is 9.1 milliseconds per $t$. It illustrates that the proposed DRL algorithm holds the promise to real-time emergency control.

### B. Result Evaluation and Comparison

To verify the effectiveness of the proposed algorithm, two indices $Q_1$ and $Q_2$ are used to evaluate the overall performance of load shedding. $Q_1$ is the success rate for the UVLS that satisfies the TVRC standard, i.e., $\alpha = 1$ in (10), in all test cases; further, $Q_2$ is the percentage of remaining loads accounting for the total original controllable loads in each test. They are expressed below.

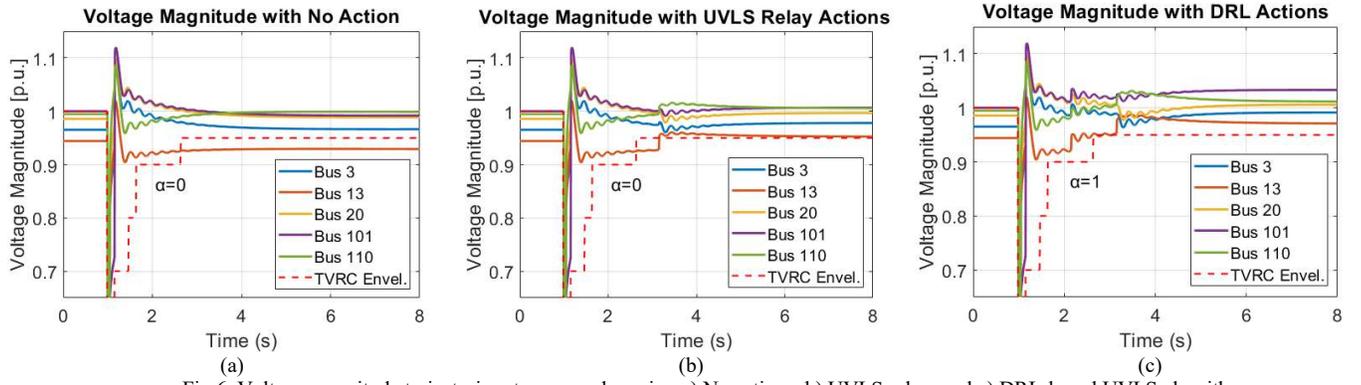

Fig.6. Voltage magnitude trajectories at some nodes using a) No actions, b) UVLS relay, and c) DRL-based UVLS algorithm

TABLE II
OVERALL PERFORMANCE EVALUATION OF DIFFERENT METHODS

| Metrics for 2000 tests | 12-bus test system | | 68-bus test system | |
|---|---|---|---|---|
| | Proposed method | UVLS relay | Proposed method | UVLS relay |
| $Q_1$ | **98.40%** | 0.75% | **89.15%** | 0% |
| Average $Q_2$ | **88.54%** | 81.22% | **93.82%** | 85.55% |

$$Q_1 = N_{succ}/M \times 100\% \quad (15)$$
$$Q_2 = \sum_{i=1}^{n} P_{Li}(t_{end})/P_{L0} \times 100\% \quad (16)$$

where $N_{succ}$ and $M$ are the number of test cases that the voltage magnitudes satisfy the TVRC and the total number of all the test cases; $P_{Li}(t_{end})$ denotes the remaining loads on bus $i$ when the UVLS process terminates. The time-series actions with the greater $Q_1$ and $Q_2$ has better shedding performances.

Fig. 6 depicts the trajectories of voltage magnitudes during the shedding process when a fault happened at line 13-101. The nodal voltages in the proposed method recover quickly above the TVRC envelope. In contrast, the UVLS relay approach is lagging in giving shedding actions, rather than in a near real-time manner, shown in Fig.6(b), since this traditional method cannot forecast the voltage trend.

Table II compares $Q_1$ and average $Q_2$ with the baseline in 2000 online tests. The actions provided by the UVLS relay accompany a low success rate of 0.75% in the two-area test system. This implicates this traditional method in most cases cannot recover the voltages timely and meet the requirements of TVRC, even when more loads are shed than the proposed method, according to average $Q_2$ in all the 2000 tests. By using the DRL shedding strategies, the averaged remaining loads are 88.54% of the initial loads, with a success rate of 98.40%. Hence, our method provides more effective actions and sheds fewer loads. In summary, the proposed algorithm outperforms the UVLS relay in both the timeliness and efficacy of load shedding.

V. CONCLUSION

This paper proposes a novel and adaptive DRL-based UVLS algorithm for grid emergency control. We pioneer a DRL framework with the customized states and rewards for UVLS following TVRC. The historical experiences from the UVLS relay are adopted as the expert system to provide some safe actions. In contrast to the model-based methods, the proposed algorithm in a real-time manner and has the adaptivity for new operation conditions. Future work will focus on improving the learning efficiency of the DRL process for distributed emergency control in large-scale power grids.

REFERENCES

[1] R. M. Larik, M. W. Mustafa, and M. N. Aman, "A critical review of the state-of-art schemes for under voltage load shedding," *Int. Trans. Electr. Energy Syst.*, vol. 29, no. 5, p. e2828, 2019.
[2] Y. Zhang, J. Wang, and M. E. Khodayar, "Graph-Based Faulted Line Identification Using Micro-PMU Data in Distribution Systems," *IEEE Trans. Smart Grid*, vol. 11, no. 5, pp. 3982–3992, Sep. 2020.
[3] D. Lefebvre, C. Moors, and T. Van Cutsem, "Design of an undervoltage load shedding scheme for the Hydro-Quebec system," in *2003 IEEE PES General Meeting*, Jul. 2003, vol. 4, pp. 2030-2036.
[4] Z. Li, G. Yao, G. Geng, and Q. Jiang, "An Efficient Optimal Control Method for Open-Loop Transient Stability Emergency Control," *IEEE Trans. Power Syst.*, vol. 32, no. 4, pp. 2704–2713, Jul. 2017.
[5] L. Jin, R. Kumar, and N. Elia, "Model Predictive Control-Based Real-Time Power System Protection Schemes," *IEEE Trans. Power Syst.*, vol. 25, no. 2, pp. 988–998, May 2010.
[6] I. Genc, R. Diao, V. Vittal, S. Kolluri, and S. Mandal, "Decision Tree-Based Preventive and Corrective Control Applications for Dynamic Security Enhancement in Power Systems," *IEEE Trans. Power Syst.*, vol. 25, no. 3, pp. 1611–1619, Aug. 2010.
[7] Q. Li, Y. Xu, and C. Ren, "A Hierarchical Data-Driven Method for Event-based Load Shedding Against Fault-Induced Delayed Voltage Recovery in Power Systems," *IEEE Trans. Ind. Inform.*, pp. 1–1, 2020.
[8] D. Ernst, M. Glavic, and L. Wehenkel, "Power systems stability control: reinforcement learning framework," *IEEE Trans. Power Syst.*, vol. 19, no. 1, pp. 427–435, Feb. 2004.
[9] M. Glavic, R. Fonteneau, and D. Ernst, "Reinforcement Learning for Electric Power System Decision and Control: Past Considerations and Perspectives," *IFAC-Pap.*, vol. 50, no. 1, pp. 6918–6927, Jul. 2017.
[10] Q. Huang, R. Huang, W. Hao, J. Tan, and Z. Huang, "Adaptive Power System Emergency Control Using Deep Reinforcement Learning," *IEEE Trans. Smart Grid*, vol. 11, no. 2, pp. 1171–1182, Mar. 2020.
[11] J. Zhang, C. Lu, C. Fang, X. Ling, and Y. Zhang, "Load Shedding Scheme with Deep Reinforcement Learning to Improve Short-term Voltage Stability," in *2018 IEEE ISGT Asia*, May 2018, pp. 13–18.
[12] P. Kundur, N. J. Balu, and M. G. Lauby, *Power System Stability and Control*, vol. 7. New York, NY, USA: McGraw-Hill, 1994.
[13] Y. Zhang, J. Wang, and Z. Li, "Uncertainty modeling of distributed energy resources: techniques and challenges," *Curr. Sustain. Energy Rep.*, vol. 6, no. 2, pp. 42–51, 2019.
[14] PJM Transmission Planning Department, "Exelon transmission planning criteria," 2009.
[15] Patterson, Dan W. *Introduction to artificial intelligence and expert systems*. Prentice-hall of India, 1990.
[16] Y. Zhang, X. Wang, J. Wang, and Y. Zhang, "Deep Reinforcement Learning Based Volt-VAR Optimization in Smart Distribution Systems," *IEEE Trans. Smart Grid*, pp. 1–1, 2020.
[17] Power System Toolbox. J. H. Chow, [Online]. Available: https://www.ecse.rpi.edu/~chowj/